\documentclass[12pt]{iopart}
\usepackage{graphicx}
\usepackage{dcolumn}
\usepackage{bm}% bold math
\usepackage{color}
\usepackage{cite}
\usepackage{hyperref}% 
\hypersetup{
    colorlinks=true,
    linkcolor=blue,
    citecolor=blue,   
   urlcolor=blue,
   }
\newcommand{\RomanNumeralCaps}[1]
    {\MakeUppercase{\romannumeral #1}}
    
\def\bea{\begin{eqnarray}}
\def\eea{\end{eqnarray}}
\def\beq{\begin{equation}}
\def\eeq{\end{equation}}
\def\f{\frac}

\def\h{\theta}
\def\t{\tau}

\def\Pe{{\rm Pe}}

\def\r{\rho}

\def\vpl{v_\parallel}
\def\vp{v_\perp}

\def\kb{k_B}

\def\la{\langle}
\def\ra{\rangle}
\def\nn{\nonumber}

\def\S{\Sigma}

\def\d{\delta}
\def\p{\partial}
\def\l{\Pe}

\def\g{\gamma}

\def\uv{ \hat{\bm{u}}}
\def\rv{ {\mathbf r}}
\def\vv{ {\mathbf v}}

\def\d{\delta}
\def\p{\partial} 

\def\la{\langle}
\def\ra{\rangle}

\def\g{\gamma}

\begin{document}
\title[Inertial dynamics of ABPs]
{Exact moments and re-entrant transitions in the inertial dynamics of active Brownian particles}

\author{Manish Patel}
\address{Institute of Physics, Sachivalaya Marg, Bhubaneswar 751005, India}
\address{Homi Bhabha National Institute, Anushaktinagar, Mumbai 400094, India}
\ead{manish.patel@iopb.res.in}

\author{Debasish Chaudhuri}
\address{Institute of Physics, Sachivalaya Marg, Bhubaneswar 751005, India}
\address{Homi Bhabha National Institute, Anushaktinagar, Mumbai 400094, India}
\ead{debc@iopb.res.in}

\begin{abstract}
In this study, we investigate the behavior of free inertial Active Brownian Particles (ABP) in the presence of thermal noise. While finding a closed-form solution for the joint distribution of positions, orientations, and velocities using the Fokker-Planck equation is generally challenging, we utilize a Laplace transform method to obtain the exact temporal evolution of all dynamical moments in arbitrary dimensions. Our expressions in $d$ dimensions reveal that inertia significantly impacts steady-state kinetic temperature and swim pressure while leaving the late-time diffusivity unchanged. Notably, as a function of activity and inertia, the steady-state velocity distribution exhibits a remarkable re-entrant crossover from ``passive" Gaussian to ``active" non-Gaussian behaviors. We construct a corresponding ``phase diagram" using the exact expression of the $d$-dimensional kurtosis. Our analytic expressions describe steady states and offer insights into time-dependent crossovers observed in moments of velocity and displacement. Our calculations can be extended to predict up to second-order moments for run-and-tumble particles (RTP) and the active Ornstein-Uhlenbeck process (AOUP). Additionally, the kurtosis shows differences from AOUP. 
\end{abstract}

%\keywords{self-propelled, inertia, exact moments, re-entrant transitions} %Use showkeys class option if keyword
                              %display desired
\maketitle

\section{\label{sec:level1}Introduction: }
Active elements can self-propel by consuming and dissipating ambient or internal energy locally, breaking detailed balance and the equilibrium fluctuation-dissipation relation~\cite{Bechinger2016, Marchetti2013, Romanczuk2012, Ramaswamy2019}. 
Natural examples of active agents are found across multiple scales, from motor proteins, cells, and tissues to insects, birds, fishes, and animals~\cite{Astumian2002, Reimann2002, Berg1972, Niwa1994, Ginelli2015, Devereux2021, Mukundarajan2016, Rabault2019}. Although relatively less, a number of artificial active systems of micro to macro scales have been engineered over the years~\cite{Bechinger2016, Marchetti2013, Howse2007, Sanchez2012, Bricard2013, Bricard2015, Ghosh2009, Dauchot2019, Scholz2018, deblais2018boundaries, Narayan2007, Kudrolli2008, Deseigne2010, Kumar2014b, Gupta2022, Scholz2018, Farhadi2018, VanZuiden2016}.
%{\color{blue}  magnetophoretic~\cite{Ghosh2009}, quincke rollers~\cite{Bricard2013, Bricard2015}, spinners}. 
Active agents performing persistent random motion are described as active Brownian particles (ABP).  Their dynamics up to the second moment agree with other related models, e.g., the run-and-tumble particles (RTP) and active Ornstein-Uhlenbeck (AOUP) process~\cite{Cates2013, Fodor2016, Das2018a, Shee2020}. They are often assumed to perform overdamped dynamics, as, e.g., the inertial relaxation time of active colloids $\sim 100$\,ns is negligibly small with respect to the persistence time. However, inertial relaxation can be much slower for larger active elements, including insects to animals and macro-sized artificial active matter {%\color{blue} 
like active granular particles, vibrated rods, vibrobots, active spinners~\cite{Narayan2007, Kudrolli2008, Deseigne2010, Kumar2014b, Gupta2022, Scholz2018, Farhadi2018, VanZuiden2016}}, and as a result, can significantly influence the emergent dynamics~\cite{Devereux2021, Mukundarajan2016, Rabault2019, Dauchot2019, Takatori2017, Scholz2018, deblais2018boundaries, nguyen21, sandoval2020pressure, herrera2021maxwell, karmakar22, Hecht2022, te2023microscopic, khali2023does}. For example, inertia can lead to the disappearance of motility-induced phase separation~\cite{Mandal2019, Lowen2020b, Caprini2022} observed for over-damped ABPs~\cite{Fily2012, Redner2013, Caporusso2020, omar23} and taming down of the instability in active nematics~\cite{chatterjee2021inertia}. 
{%\color{blue} 
Moreover, it results in a distinctive impact on the dynamics of active phase separation~\cite{Su2021}, spatial velocity correlations~\cite{Caprini2021b}, and entropy production~\cite{Shankar2018c}. Considering inertial dynamics of orientation can increase the effective persistence~\cite{lisin22, Sprenger2023}.}
Recent experiments on vibrobots showed an inertial delay between the instantaneous orientations and velocities and bimodality in the velocity distributions~\cite{Scholz2018, Lowen2020b} captured by perturbative calculations~\cite{herrera2021maxwell}.
Nevertheless, even with substantial advancements in active matter research, a thorough understanding of the dynamics of non-interacting inertial ABPs still remains elusive.

In this paper, we consider the motion of non-interacting ABPs in the presence of translational inertia. Using a Laplace transform approach, we show that the Fokker-Planck equation governing the dynamics can be utilized to evaluate the exact time dependence of all moments of dynamical variables in arbitrary $d$-dimensions. This is the first main contribution of this paper. {%\color{blue} 
Offering a unified approach for computing the precise time evolution of all conceivable dynamical variables in arbitrary dimensions, our method represents a noteworthy analytical advancement in investigating inertial effects in active matter.} The technique was originally proposed back in 1952 to study equilibrium properties of worm-like chains~\cite{Hermans1952a} and was recently put in use to study the dynamics of overdamped ABPs in Ref.~\cite{Shee2020, Chaudhuri2020, Shee2022, Shee2022a}.  

We present exact expressions of several dynamical moments, which agree with the direct numerical integration results. The system eventually reaches a steady state in velocity but not displacement, with mean-squared displacement undergoing dynamic crossovers before reaching asymptotic diffusion with inertia-independent diffusivity.  
In contrast, the kinetic temperature and swim pressure depend on inertia; with inertia, while the kinetic temperature increases to saturate eventually, the swim pressure decreases monotonically to vanish in the infinite mass limit. 

The most striking result is the occurrence of a re-entrant crossover between active and passive states as inertia varies, quantified through deviations from the equilibrium-like Gaussian velocity distribution. Analytically, this is captured by the $d$-dimensional kurtosis of velocity, which shows a non-monotonic variation with inertia and a decrease to negative values followed by a saturation with increasing activity. We use the results to obtain a ``phase diagram" in the activity-inertia plane. This is the second main contribution of this study. Note that in this single-particle system, there is no genuine phase transition; instead, it represents a crossover between active and passive behaviors.
Further, we propose approximate analytic forms of probability distributions of velocity that show reasonable agreement with numerical results at small and large inertia limits and can qualitatively capture the re-entrant active-passive crossovers.  
Our calculations for ABP can be extended to predict up to second-order moments for RTPs and the AOUPs. Additionally, the non-zero kurtosis for intermediate inertia shows differences from AOUP.

The paper is structured as follows: In Section~\ref{sec_model}, we introduce the model and Langevin dynamics. The subsequent section outlines a Laplace-transform-based approach to compute dynamic moments in any dimension. Sections~\ref{sec_vel} and \ref{sec_disp} present detailed results related to velocity and displacement moments. Section~\ref{sec_vel} delves into discussions about kinetic temperature, the ``phase diagram", and velocity distributions, while Section~\ref{sec_disp} explores topics like active diffusivity and swim pressure. {%\color{blue} 
Using appropriate mapping, we extend the results to the related RTP and AOUP systems in Section~\ref{sec_discuss}.} Finally, in Section~\ref{sec_conc}, we provide a conclusion summarizing our key findings.

%%=---------------- Model---------------------------------%%
\section{Model}
\label{sec_model}
The underdamped motion of ABPs in $d$-dimensional space moving with a constant active speed $v_a$ in the orientation $\uv = (u_1,u_2,...,u_d)$ is described by its position $\rv' = (r_1',r_2',...,r_d')$ and velocity $\vv' = (v_1',v_2',...,v_d')$ evolving with time $t'$. We consider the dynamics in the presence of translational and orientational Brownian noise with diffusivities $D$ and $D_r$, respectively. 
The time and length scales are set by $\t_r= D_r^{-1}$ and $\ell=\sqrt{D/D_r}$, to give the unit of velocity $\bar v = \sqrt{D D_r}$. We use dimensionless position $\rv = \rv'/\ell$, time $t=t'/\t_r$, and velocity $\vv=\vv'/\bar v$  to express the Langevin equation of motion in $d$ dimensions in the following Ito form~\cite{Ito1975,VandenBerg1985,Mijatovic2020} 
\bea
  \mathrm{d}r_i &=& v_i \mathrm{d}t,\label{eom:disp_abp_speed_fluct}\\
M \mathrm{d}v_i &=& -(v_i - \l \, u_i) \mathrm{d}t + \sqrt{2}\, \mathrm{d}B_i(t),\label{eom:speed_abp_speed_fluct}\\
\mathrm{d} u_i&=& (\d_{ij}-u_i u_j)\, \sqrt{2}\, \mathrm{d}B_{j} (t) -(d-1) u_i \mathrm{d}t.\label{eom:rot_Ito_speed_fluct}
\eea
where we used the dimensionless mass $M=\t/\t_r$ with inertial relaxation time  $\t = m/\g$, $m$ denoting the translational inertia and $\g$ the viscous damping, and P{\'e}clet number $\l=v_a/\sqrt{D D_r}$. The Gaussian noise in translation and rotation are uncorrelated, and their components obey $\la \mathrm{d}B_i\ra=0$ and $\la \mathrm{d}B_i \mathrm{d}B_j\ra = \d_{ij}\,\mathrm{d}t$.  The first term in Eq. (\ref{eom:rot_Ito_speed_fluct}) denotes a projection operator for the noise onto a $(d-1)$-dimensional surface perpendicular to $\uv$. The second term ensures that the unit vector remains normalized after displacement. The equations can be directly integrated numerically using the Euler-Maruyama scheme.

%%----------------- Calculations: FP eqn and moments----------------- %%

\section{Calculation of moments from the Fokker-Planck equation}
\label{sec_FP}
Noting that the heading direction $\uv$ undergoes an orientational diffusion on a $(d-1)$-dimensional hypersurface and $\vv$ performs a drift-diffusion, 
the probability distribution $P(\rv,\vv,\uv,t)$ evolves following the Fokker-Planck equation 
%%----------------  Fokker-Planck Equation -------------------------------%%
\bea
    \p_t P = -\nabla \cdot(\vv\, P) - (1/M)\nabla_v \cdot [(\l\, \uv - \vv)P] + ( 1/M^2) \nabla_v^2 P + \nabla_u^2 P
    \eea
where $\nabla$ and $\nabla_v$ denote the gradient operator in $d$-dimensional position and velocity space, respectively. 
The Laplacian $\nabla_u^2$ in $(d-1)$-dimensional orientation space can be expressed in terms of Cartesian coordinates ${\bf y}$ as  
$\nabla_u^2 = y^2 \sum_{i=1}^d \p_{yi}^2-[y^2\p_y^2 +(d-1)y\p_y]$ 
by defining $u_i = y_i/y$ with $y = |\bf{y}|$.
Using a Laplace transform $\tilde{P}(\rv,\vv,\uv,s) = \int_{0}^{\infty} dt\, e^{-st}\, P(\rv,\vv,\uv,t)$, the Fokker-Planck equation can be expressed as
\bea
\fl 
- P(\rv,\vv,\uv,0) + s\tilde{P}(\rv,\vv,\uv,s) 
= -\nabla\cdot(\vv \tilde{P})- \f{1}{M} \nabla_v \cdot [(\l \,\uv - \vv)\tilde{P}] + \f{1}{M^{2}}\nabla_v^2 \tilde{P} + \nabla_u^2 \tilde{P} \nn
\eea
%
%%----------------  Observable Equation -------------------------------%%
Therefore, the mean of any observable $\la \psi \ra_s = \int d\rv \, d\vv \,  d\uv\, \psi(\rv,\vv,\uv) \tilde{P}(\rv, \vv, \uv,s)$ satisfies the moment equation
\bea 
\label{observable}
\fl
     -\la \psi \ra_0 + s \la \psi \ra_s &= \la \vv \cdot \nabla \psi \ra_s + \f{\l}{M} \la \uv \cdot \nabla_v \psi \ra_s - \f{1}{M} \la \vv \cdot \nabla_v \psi \ra_s  %\nn\\
    %&
    + \f{1}{M^2} \la \nabla_v^2 \psi \ra_s +  \la \nabla_u^2 \psi \ra_s,
\eea
where $\la \psi \ra_0 = \int d\rv\, d\vv \, d\uv\, \psi(\rv, \vv,\uv) P (\rv, \vv, \uv,0)$ is given by the initial condition, which without any loss of generality, can be expressed as  $P(\rv,\vv,\uv,0) = \d(\rv) \d(\vv - \vv_0) \d(\uv - \uv_0)$. 
Equation\,(\ref{observable}) can be used to determine any dynamical moment of interest in arbitrary \textit{d}-dimensions as a function of time by performing an inverse Laplace transform. In the following, we denote the steady-state values $\lim_{t\to\infty} \la \psi\ra(t) \equiv \la \psi\ra_{st} = \lim_{s\to 0} s\la \psi \ra_s$.  Our theoretical approach, {%\color{blue} 
offering a unified method for computing the precise time evolution of all dynamical variables in arbitrary dimensions,} differs from the recent works on inertial ABPs \cite{Scholz2018,herrera2021maxwell,lisin22, Sprenger2023}. {%\color{blue} 
Moreover, we present approximate analytic expressions for velocity distributions parallel and perpendicular to the heading direction in Sec.~\ref{sec_prob}. }
   
\section{Velocity moments, steady states, and ``phase transitions"}
\label{sec_vel}
We first employ the above equation to obtain the time evolution of several velocity moments, {%\color{blue} 
including mean, variance, quartic moment, and lag function} towards the steady state. The dynamics of inertial passive particles lead to 
the equilibrium Maxwell-Boltzmann distribution of velocity. Activity changes this behavior~\cite{Scholz2018, Lowen2020b}. Here, we calculate up to fourth order cumulant of velocity to identify such deviations and hence determine a ``phase diagram" describing departures from the equilibrium Gaussian characteristic.  
  
\subsection{Mean velocity}
Using $\psi = \vv$ in equation\,(\ref{observable}) we get the velocity in the Laplace space 
$\la \vv \ra_s = \frac{\vv_0 + \l \la \uv \ra_s/M}{s+ 1/M}$
in terms of $\la \uv \ra_s$.
Again, using $\psi=\uv$, equation\,(\ref{observable}) gives
$\la \uv \ra_s = \uv_0/[s+(d-1)]$.  Together, these two expressions lead to 
\begin{equation} \label{v_lap}
    \la \vv \ra_s = \frac{\vv_0}{s + M^{-1}} + \frac{\l \, \uv_0}{M(s+d-1)(s + M^{-1})}.
\end{equation}
Expressions of dynamical variables with $s$ dependence $\prod_i (s+s_i)^{-1}$ lead to sum over exponentials in time after inverse Laplace transform. Thus, $\la \uv \ra(t) = \uv_0 e^{-(d-1)t}$ and  
\bea
\la \vv \ra(t) = \vv_0 e^{-t/M}+\f{\l \, \uv_0}{(d-1)M-1}\left( e^{-t/M}-e^{-(d-1)t}\right).
\eea
The mean velocity vanishes in the steady state, $\la \vv \ra_{st}=0$.

\begin{figure}[t]
    \includegraphics[scale = 0.6]{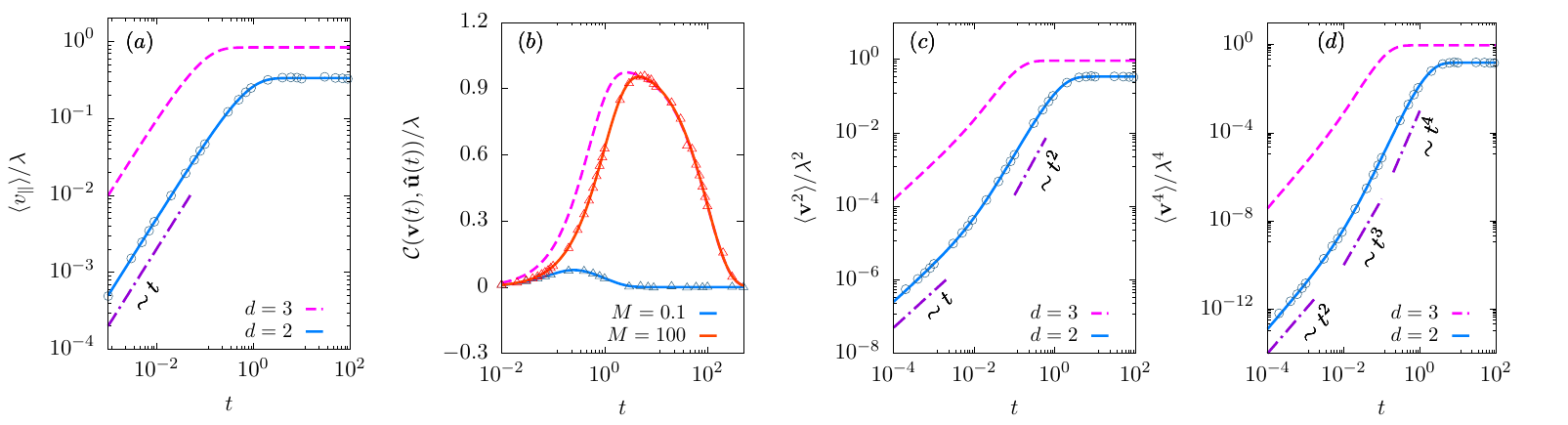}
    \caption{Time evolution of velocity moments $\la \vpl\ra$~($a$), $\la \vv^2\ra$~($c$), $\la \vv^4\ra$~($d$) in $d=2,3$ with initial condition $\vv_0={\bf 0}$. In ($a$), ($c$), ($d$), we use $\l=20$  and $M=2$ for $d=2$ and $M=0.1$ in $d=3$. In ($b$), the delay function ${\cal{C}}(\vv(t),\uv(t))$ is shown at $\l=100$ in 2d for two $M$ values (solid lines) and for $M=100$ in $d=3$ (dashed line). The lines in $(a)$, $(b)$, $(c)$ and $(d)$ are the plots of equations\,(\ref{eq_uvt}), (\ref{eq_delay}), (\ref{eq_v2t}), and (\ref{eq_v4t}) respectively and the points denote simulation result in 2d. We indicate crossovers between different scaling regimes in ($c$) and $(d)$. In ($c$), crossover between $\sim t$ to $\sim t^2$ appears at $t_I = 0.01$. ($d$) shows crossovers from $\sim t^2$ to $\sim t^3$ at time $t_I  = 0.005$ and $\sim t^3$ to $\sim t^4$ at $t_{II}=0.04$ for $d=2$. 
     } 
     \label{fig_vv}
\end{figure}

In the presence of inertia, the orientation of the velocity vector $\vv$ can be different from the heading direction $\uv$. {%\color{blue} 
Orientation fluctuation of the heading direction over the inertial relaxation time leads to this difference. In the limit of vanishing inertia, one expects $\la v_\parallel \ra = \Pe$.} Let us define instantaneous velocity components parallel and perpendicular to the heading direction, $\vv_{\parallel}= (\vv \cdot \uv) \uv$, $\vv_\perp = \vv-\vv_\parallel$. 
Using $\psi = v_{\parallel}=\uv \cdot \vv$ in equation(\ref{observable}) 
we get
\bea \label{eq_uvs}
\la v_{\parallel} \ra_s \equiv \frac{(\uv_0 \cdot \vv_0)+\l/(sM)}{s+1/M+(d-1)}
\eea
Its inverse Laplace transform yields,
\bea \label{eq_uvt} \fl
\la v_{\parallel} \ra (t) = \frac{\l}{(d-1)M+1} +\left((\uv_0 \cdot \vv_0)-\frac{\l}{(d-1)M+1} \right)e^{-((d-1)+M^{-1})t}
\eea
Thus $\la v_\parallel\ra$ starts from $\uv_0\cdot \vv_0$ at $t=0$ to asymptotically reach the steady-state value $\frac{\l}{(d-1)M+1}$~(see Fig.\ref{fig_vv}($a$)\,). {%\color{blue} 
As expected,} in the limit of vanishing inertia, $\la v_{\parallel} \ra = \l$, the velocity vector always remains oriented along the heading direction. The larger the inertia, the bigger the deviation from the heading direction~(also see Fig.~\ref{fig_v_initial} in the Appendix). By symmetry $\la \vv_\perp\ra=0$. 

%%----------------  Moments Calculation -------------------------------%%
%%---------------- Figure - 1 -------------------------------%%

\subsection{Lag function}
While analyzing the motion of inertial vibrobots, a lag function was used in Ref.~\cite{Scholz2018}, 
${\cal{C}} ( \vv(t), \uv (t)) = \la \vv(t) \cdot \uv(0) \ra - \la  \uv(t) \cdot \vv(0) \ra$, 
which quantifies the average difference between the projection of velocity to initial heading direction and the projection of heading direction to the initial velocity. In equilibrium, time-reversal symmetry ensures ${\cal{C}} ( \vv(t), \uv (t)) =0$. 
We obtain the following closed-form expression for inertial ABPs, 
\bea \label{eq_delay}
  {\cal{C}} ( \vv(t), \uv (t)) = \frac{\l}{1-[(d-1)M]^{-1}} \left( e^{-t/M} - e^{-(d-1)t} \right),
\eea
using $\vv(0) = \l \uv_0$.
In the absence of inertia, ${\cal{C}} ( \vv(t), \uv (t))=0$. In the inertial system, ${\cal{C}} ( \vv(t), \uv (t))$ starts from zero at $t=0$ to returns to zero as $t \to \infty$ after reaching a maximum at an intermediate time $t_m = \f{M}{M(d-1)-1}\ln[(d-1)M]$. Plots of this expression in $d=2,3$ are shown in Fig.\ref{fig_vv}($b$), with a comparison against numerical simulations performed in two dimensions. A similar non-monotonic variation was observed in vibrobots~\cite{Scholz2018}.
 
\subsection{Mean squared velocity}
Using $\psi = \vv^2$ in equation~(\ref{observable}) and $\la \psi \ra(0) = \vv_0^2$, $\vv \cdot \nabla \psi = 0$, $\la \vv \cdot \nabla_v \psi \ra_s = 2 \la \vv^2 \ra_s$, $\la \uv \cdot \nabla_v \psi \ra_s = 2 \la \uv \cdot \vv \ra_s $, $\la \nabla_v^2 \ra_s 
= 2d \la 1 \ra_s 
= 2d/s $~\footnote{As, $\la 1 \ra_s  = \int d\rv d\vv d \uv\, \tilde{P} = \int d\rv d\vv d\uv \int_{0}^{\infty} dt\, e^{-st} P = \int_{0}^{\infty} dt\, e^{-st} \{d \rv d \vv d \uv P\} = \int_{0}^{\infty} dt\, e^{-st} = 1/s$} and $\nabla^2 \psi = 0$, we get  
$
    \la \vv^2 \ra_s = \frac{1}{s+2/M}\left[ \vv_0^2 + \frac{2\l\la v_{\parallel} \ra_s}{M} + \frac{2d}{sM^2} \right]. 
$    
Using equation(\ref{eq_uvs}), we get
\bea\label{finalv2s}
\la \vv^2 \ra_s &= \frac{1}{s+2/M}\left[ \vv_0^2 +  \frac{2 \l(\uv_0 \cdot \vv_0 + \l/(sM))}{M(s+1/M +(d-1))} %\right.  \nn\\
 + \frac{2d}{sM^2}  \right]. 
\eea
The inverse Laplace transform gives
 \bea \label{eq_v2t}
 \fl
    \la {\vv}^2 \ra ({t}) = \frac{\l^2}{(d-1) M+1}+\frac{d}{M}+ \left( \left( \vv_0^2 - \frac{d}{M}\right) + \frac{2(\uv_0 \cdot \vv_0) \l - \l^2}{(d-1)M-1} \right) e^{-2{t}/M}  \nn\\
     - \frac{2 \Pe   [\,(\uv_0 \cdot \vv_0) \{(d-1) M+1\}-\Pe \,]}{(d-1)^2 M^2-1}e^{-[(d-1)+M^{-1}]{t}}
 \eea
We plot this expression in Fig.~\ref{fig_vv}($c$) for $d=2,3$ and show a comparison with numerical simulations in 2d. To analyze the observed dynamical crossovers from $\sim t$ to $\sim t^2$ growth before reaching the steady state, we expand $\la \vv^2 \ra (t)$ around $t=0$ using $\vv_0 = {\bf 0}$, 
\bea \label{eq_v2series} \fl
 \la \vv^2 \ra (t) = \frac{2d}{M^2}t + \frac{(\l^2M-2d)}{M^3}t^2 + \frac{d \left(4-\l^2 M^2\right)+\l^2 (M-3) M}{3M^4} t^4 + {\cal{O}} (t^5).
\eea
The expansion shows a $\la \vv^2 \ra \sim t$ scaling at a short time before a crossover to $\la \vv^2 \ra \sim t^2$ at $t_{\RomanNumeralCaps{1}} = 2 d M/(\l^2 M-2 d)$. For crossovers starting from a different initial condition, see \ref{sec_app_vel}. 
In the asymptotic limit of long time $t \gg \f{M}{2}$ and $t \gg \f{M}{M(d-1)+1}$, the mean-squared velocity reaches the steady state value
\bea
\la \vv^2 \ra_{st}= \frac{\l^2}{(d-1) M+1}+\frac{d}{M}. 
\eea
Given that $\la \vv \ra_{st}=0$, the velocity fluctuation $\la \d \vv^2\ra_{st}=\la \vv^2\ra_{st}$.

\subsection{Effective temperature and violation of equilibrium fluctuation-dissipation relation}
The expression of mean squared velocity can be used to obtain the steady-state kinetic temperature of the system
\bea
k_B T_{\rm kin} = \f{M \la \vv^2\ra_{st}}{d} = 1 + \f{\l^2}{d}\frac{M}{(d-1) M+1}. 
\label{eq_Tkin}
\eea
Remarkably, unlike in equilibrium, the kinetic temperature is a function of inertia. In the over-damped limit $k_B T_{\rm kin}=1$. The kinetic temperature increases with $M$ linearly at a small $M$ to saturate to $1+\l^2/[d(d-1)]$ at large $M$. {%\color{blue} 
The kinetic temperature in the coexistence of phase separating ABPs~\cite{Mandal2019} and in active Ornstein-Uhlenbeck particles in two dimensions~\cite{Caprini2021} were calculated before. With a careful mapping of the active force and persistence time of ABPs to that of Ornstein-Uhlenbeck particles, the above prediction of kinetic temperature agrees with that of Ref.\cite{Caprini2021} in $d=2$~(see Sec.\ref{sec_discuss}).}  

At equilibrium ${\cal I} = D-\mu \kb T = 0$, due to the equilibrium fluctuation-dissipation relation. Using the expression of effective diffusivity $D_{\rm eff}=1+\l^2/[d(d-1)]$ derived later in equation\,(\ref{eq_Deff}) in Sec.~\ref{sec_disp}, we obtain the dimensionless form of the violation of equilibrium fluctuation-dissipation \bea\label{eq_nonequilibrium}
    {\cal{I}} = D_{\rm eff} - k_B T_{\rm kin}
    =\f{\l^2}{d(d-1)}\left[1 - \f{(d-1)M}{(d-1)M+1}\right]. 
 \eea
Clearly, the violation is maximum in the overdamped limit of $M=0$. It decreases linearly with $M$ at small $M$, to vanish asymptotically as $M \to \infty$. Thus, with increasing $M$, ABPs can return to equilibrium-like behavior.

\subsection{Fluctuation of velocity component in the heading direction}
As we have shown before, 
\bea
\la \vpl\ra_{st}= \frac{\l}{(d-1)M+1}, 
\label{eq_vplst}
\eea
the steady state value of $\la \vpl\ra_{st}$ decreases from $\l$ to $0$ monotonically, with increasing mass~(Fig.~\ref{fig_vpl_st}($a$)\, in \ref{sec_app_vel}). 
Here we calculate the variance $\la \d v_{\parallel}^2 \ra = \la v_{\parallel}^2 \ra - \la v_{\parallel} \ra^2$.
Using  $\psi = \vpl^2 \equiv (\uv \cdot \vv)^2 $ in equation(\ref{observable}) 
\bea 
    (s+2/M + 2d) \la \vpl ^2 \ra_s = (\uv_0 \cdot \vv_0)^2 + \frac{2\l}{M} \la \vpl \ra_s + \frac{2}{sM^2} +2 \la \vv^2 \ra_s \nn
\eea
While it is straightforward to calculate the full time-evolution of $\la \vpl^2\ra(t)$ performing an inverse Laplace transform of the above expression, 
we show the steady-state expressions
\bea \label{eq_vp}
 \la v_{\parallel}^2 \ra_{st} &=& \frac{\l^2 (M+1)}{((d-1) M+1) (d M+1)}+\frac{1}{M} \\
\label{eq_dv2p}
\la \d v_{\parallel}^2 \ra_{st} 
&=& \frac{(d-1) \Pe ^2 M^2}{((d-1) M+1)^2 (d M+1)}+\frac{1}{M}
 \eea
 In the limit of vanishing inertia, the above expression gives $M \la \d v_\parallel^2\ra_{st}=1$ and matches with the overdamped limit of $k_B T_{kin}=1$. In the other limit of $M\to \infty$, $M \la \d v_\parallel^2\ra_{st} = 1+\f{\l^2}{d(d-1)}$, again agrees with the $M\to \infty$ limit of $k_B T_{kin}$. It is important to note that, in general, for intermediate values of $M$,  $M \la \d v_\parallel^2\ra_{st} \neq \kb T_{kin}$. The $\la \d v_\parallel^2\ra_{st} \sim 1/M$ scalings in the small and large $M$ limits are observable in Fig.~\ref{fig_vpl_st}($b$) of \ref{sec_app_vel}.
 Further, using the definition $\la \vp^2\ra_{st} = \la \vv^2\ra_{st} - \la \vpl^2\ra_{st}$  we get
 \bea
  \label{eq_vprp}
 \la \vp^2 \ra_{st} = (d-1)\left[ \f{\l^2 M}{((d-1)M+1)(d M +1)} +\f{1}{M} \right] 
 \eea
 and $\la \d \vp^2\ra_{st}=\la \vp^2\ra_{st}$ as $\la \vp \ra=0$. In the overdamped limit $M \la \vp^2\ra = (d-1)$ and in the limit of $M\to \infty$ we find $M \la \vp^2\ra = (d-1)[1+\f{\l^2}{d(d-1)}]$. Fig.~\ref{fig_vpl_st}($c$) in \ref{sec_app_vel} shows the $M^{-1}$ scaling of $\la \vp^2\ra$ in the small and large $M$ limits.

\subsection{Fourth moment of velocity}

The calculation of $\la \vv^4 \ra_s$ proceeds as before using equation(\ref{observable})  and involves the following steps, 
\bea 
\label{v4s}
\fl
    (s+4/M) \la \vv^4 \ra_s = \vv_0^4 + \frac{4\l}{M} \la (\uv \cdot \vv)\vv^2 \ra_s + \frac{4(d+2)}{M^2} \la \vv^2 \ra_s,   \\
\label{uvv2s}
\fl
    (s+3/M +(d-1)) \la (\uv \cdot \vv)\vv^2 \ra_s = (\uv_0 \cdot \vv_0)\vv_0^2 + \frac{\l}{M} \la \vv^2 \ra_s + \frac{2\l}{M} \la (\uv \cdot \vv)^2 \ra_s 
     + \frac{2(d+2)}{M^2} \la \uv \cdot \vv \ra_s, \nn\\
\eea
where $ \la \uv \cdot \vv \ra_s$, $\la \vv^2 \ra_s$ and $\la (\uv \cdot \vv)^2\ra_s$ are already calculated in equations (\ref{eq_uvs}), (\ref{finalv2s}) and (\ref{eq_vp}) respectively. 
Using them, we get the exact form of $\la \vv^4 \ra_s$ as shown in \ref{sec_v4}. The inverse Laplace transform
gives the time-dependence $\la \vv^4 \ra (t)$ shown in equation(\ref{eq_v4t}) of Appendix. 
Here, we present the steady-state result, 
\bea
\fl
 \la {\vv}^4 \ra_{st} = \frac{d (d+2)}{M^2}+\frac{2 (d+2) \l^2}{((d-1) M+1)M}  +\frac{\l^4 ((d+2) M+3)}{((d-1) M+1) ((d-1) M+3) (d M+1)}
\eea
The short-time behavior with initial condition $\vv_0={\bf 0}$ can be analyzed in terms of a series expansion of $\la \vv^4 \ra (t)$ around 
$t = 0$ yielding,
\bea 
\la \vv^4 \ra (t) = \frac{4d(d+2)}{M^4}t^2 + {\cal A}_1 t^3 + {\cal B}_1 t^4 + {\cal{O}} (t^5), \nn
\eea
where ${\cal A}_1=4 [ (d+2)( \l^2 M -2d )] /M^5$, and $3 {\cal B}_1 M^6=[\l^2 M^2 \{-4 d (d+1)+3 \l^2+8\}-24 (d+2) \l^2 M+28 d (d+2)]$. 
As shown in Fig.~\ref{fig_vv}($d$), $\la \vv^4 \ra (t)$  undergoes a crossover from $\sim t^2$ scaling at short times to $\sim t^3$ scaling  near  $t_I = \frac{4d(d+2)}{M^4}/{\cal A}_1 =dM/(\l^2 M -2d)$.  Further, it crosses over to a $\sim t^4$ scaling at $t_{II} = {\cal A}_1/{\cal B}_1$ before saturating to a steady state. Time-dependence of velocity moments before reaching a steady state depends on the initial condition, a point illustrated further in \ref{sec_app_vel}.

\begin{figure}[t]
\centering
    \includegraphics[scale = 0.14]{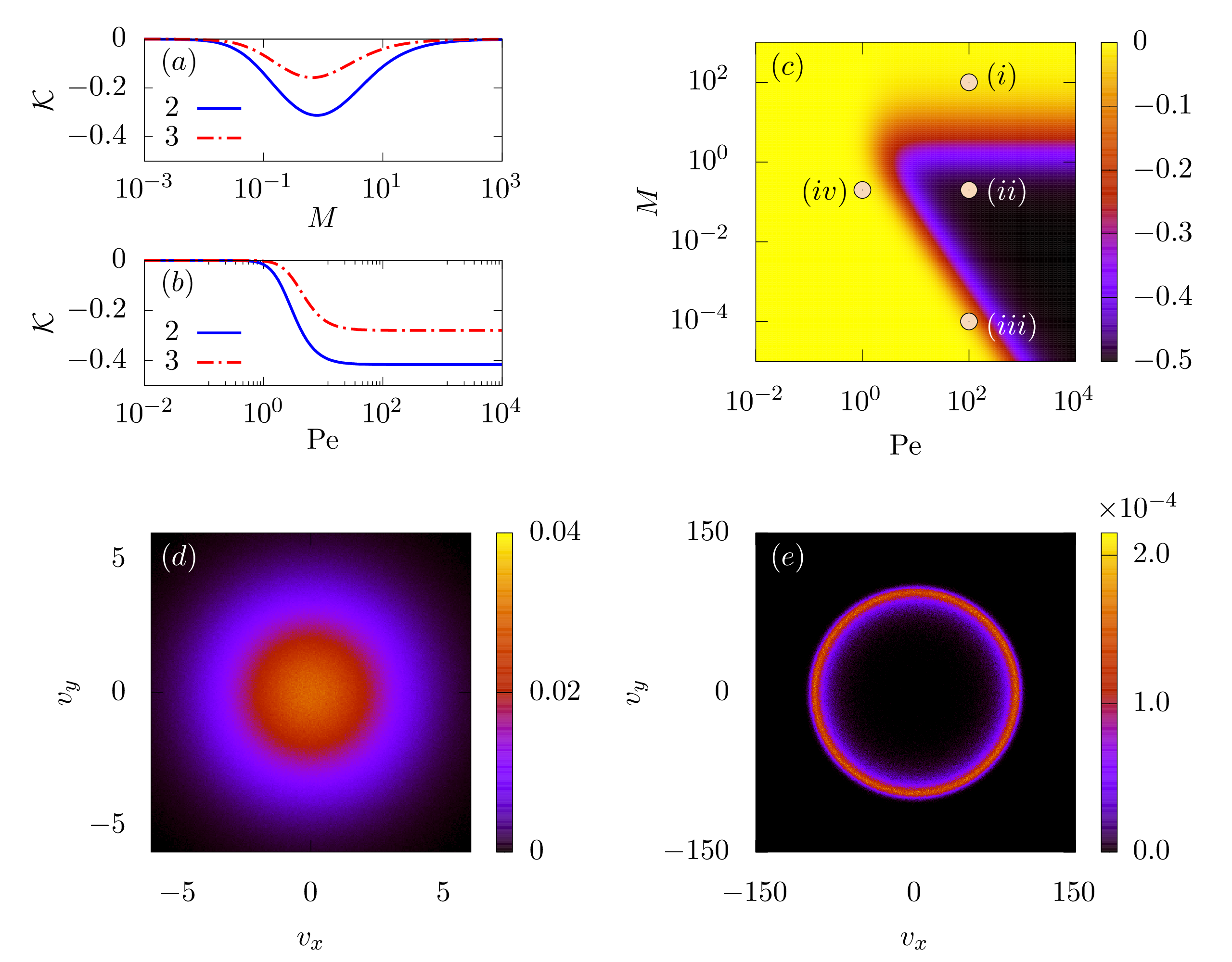}
    \caption{The steady-state kurtosis $\cal{K}$ as a function of $M$ at $\l=10$ ($a$) and as a function of activity $\l$ at $M=1$ ($b$). The solid (dashed) line shows the plot in $d=2$ ($d=3$). ($c$)~{``Phase diagram"}: A heat map of steady-state kurtosis as a function of activity $\l$ and mass $M$ in $d = 2$. Passive and active states are represented by light (yellow) and dark (black) regions, respectively. 
     ($d$)~Heat maps of $P(v_x,v_y)$ in the passive state identified by point ($iv$) in the ``phase diagram" ($c$) 
    and ($e$)~$P(v_x,v_y)$ in the active state identified by point ($ii$) in ($c$).   }
    \label{fig_vdistribution}
\end{figure}

%%----------------  Kurtosis-------------------------------%%

\subsection{Kurtosis: deviations from the Gaussian process}

The fourth moment $\la \vv^4 \ra$ for a general Gaussian process with $\la \vv \ra=0$ in terms of lower order moments can be expressed as~\cite{Shee2020}
\bea
    \mu_4 := \left(1+\frac{2}{d}\right)\la \vv^2\ra^2
\eea
The departure from Gaussian behavior can be evaluated using the kurtosis
 \bea
 {\cal K} = \frac{\la \vv^4 \ra}{\mu_4} - 1. 
 \eea
At steady-state, it has the form
\bea
  {\cal K} =   \frac{-2 \Pe ^4 M^2 ((4 d-1)M+3)}{(d+2) ((d-1) M+3) (d M+1) \left(d^2 M-d M+d+\Pe ^2 M\right)^2}.  
  \label{eq_k}  
\eea
Thus, kurtosis shows a non-monotonic variation with $M$~(Fig.\ref{fig_vdistribution}(a)\,). It vanishes in the two limits of small and large inertia --  in the limit of $M \to 0$ as 
\bea
{\cal K} \approx -\frac{2 \l^4}{3 d^2 (d+2)}\, M^2
\label{eq_K_od}
\eea
and in the limit of $M \to \infty$ as ${\cal{K}} \sim  M^{-1} $,
indicating the two {\em passive} states with asymptotically Gaussian distributions of velocity. For intermediate $M$ values, ${\cal K}$ becomes negative and reaches a minimum. This suggests a re-entrance transition from passive to active to passive behavior with the increase in inertia. 

On the other hand, with activity $\l$, kurtosis decreases monotonically to saturate~(Fig.\ref{fig_vdistribution}($b$)\,). For $\l \to 0$, it vanishes as ${\cal{K}} \sim \l^4$ . In the limit of large activity $\l \to \infty$, it saturates to 
\bea
{\cal {K}} = -\frac{2(4dM-M+3)}{(d+2)(dM+1)((d-1)M+3)}. 
\eea
The large amplitude of kurtosis characterizes a strongly non-Gaussian velocity distribution denoting an active state. As shown in Fig.\ref{fig_vdistribution}($d$), the velocity distribution in two dimensions (2d) corresponding to the active state obtained from numerical simulations displays a distinct ring shape.

\subsection{``Phase diagram" describing active-passive crossovers}  
  
A heat map of equation~(\ref{eq_k}) in Fig.\ref{fig_vdistribution}($c$) produces a ``phase diagram" identifying crossovers between active (dark, large amplitude of ${\cal K}$) and passive (light, ${\cal K}\approx 0$) regions in 2d. It clearly shows a re-entrant crossover from passive (Gaussian) to active (non-Gaussian) to passive (Gaussian) with increasing inertia at a constant activity, e.g., near $\l \approx 100$. The ``phase diagram" in $d=3$ also shows similar behavior~(figure not shown).  
  
  The re-entrant crossover can be understood using the following heuristic argument. The velocity distribution becomes Gaussian if the persistence of the heading direction is lost over the inertial relaxation time, i.e., $\t \gg \t_r$,  which is equivalent to $M\gg 1$. {%\color{blue} 
  As a result, the active fluctuations can be approximated as additional `thermal noise.'} The ``phase diagram" shows the actual boundary is near $M=1$. At $M<1$, i.e., $\t < \t_r$, persistence dominates; therefore, the velocity distribution shows maximum on a  $(d-1)$-dimensional hypersphere of radius $v=\l$, with the spherical shape emerging due to isotropy in the initial orientation. Also, the thermal noise thickens the hypersurface shell by $\la \d v^2 \ra \approx d/M$. The $v=\l$ peak of the distribution gets unrecognizable when the velocity fluctuation gets larger than the ring size, $\sqrt{\la \d v^2 \ra} \gg \l$, leading to the criterion $M <\l^{-2}$ for a re-entrance to the passive Gaussian regime. This criterion is consistent with equation~(\ref{eq_K_od}). 

\subsection{Probability distributions of velocity in 2d}
\label{sec_prob}

\begin{figure} 
    \centering
    \includegraphics[scale=0.7]{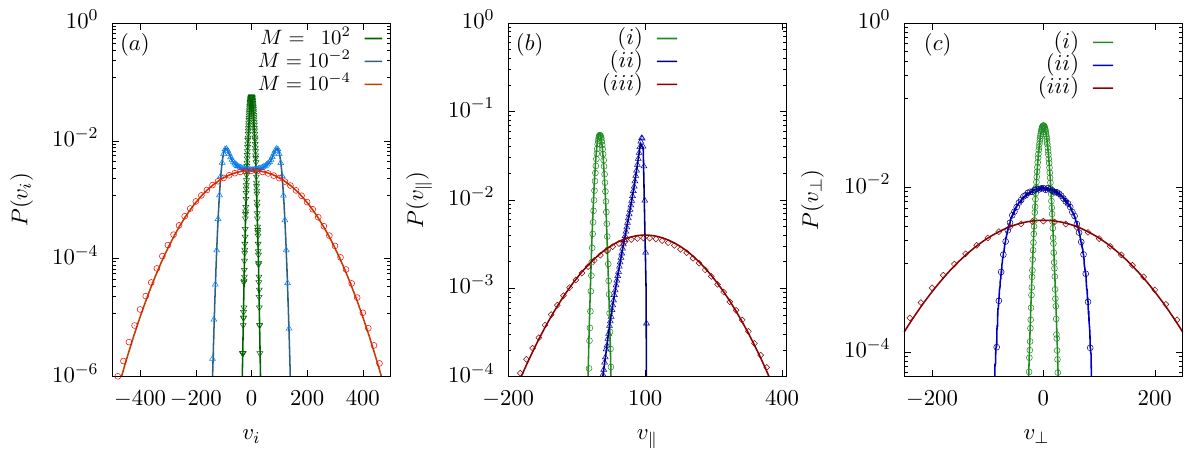} %{pvx.pdf}
    \caption{Velocity distributions in 2d at $\l = 100$: 
    ($a$)~Probability distributions of components of velocity $v_i$ with $i=x,y$. Lines denote analytic expressions, and points denote simulation results. The small $M$ Gaussian to non-Gaussian crossover is captured by equation~(\ref{eq_prob1}) showing good agreement with simulation results at $M = 10^{-4}$ and $M = 10^{-2}$. The large $M=10^2$ result from simulations agrees with equation~(\ref{eq_prob2}).  
    Probability distributions of velocity components parallel $P(\vpl)$ and perpendicular $P(\vp)$ to the heading direction are shown in ($b$) and ($c$). 
  The distributions denoted by ($i$), ($ii$), and ($iii$) in the legends correspond to the points marked in Fig.~\ref{fig_vdistribution}($b$) at fixed $\l = 100$ and $M = 100$, \,0.2, \,$10^{-4}$, respectively. The lines at $M=10^{-4}$~($iii$)  are plots of equation~(\ref{eq_vsm}) and that at  $M=100$~($i$) are plots of equation~(\ref{eq_vlm}). The lines at intermediate $M=0.2$~($ii$) are guides to the eye. 
    }
    \label{fig_pvx}
\end{figure}

While finding a closed-form solution for the joint distribution of positions, orientations, and velocities using the Fokker-Planck equation is generally challenging, it can be analyzed in the two limits of small and large $M$, particularly for the marginal distributions of velocities. {%\color{blue} 
The velocity fluctuations are governed by the kinetic temperature $\kb T_{\rm kin}$ in Eq.(\ref{eq_Tkin}).} For $M \ll 1$, i.e., $\t_r \gg \t$, the heading direction can be assumed to be fixed at some angle $\h$. This regime is equivalent to the dynamics of a particle under external force.  {%\color{blue} 
Moreover, $\kb T_{\rm kin}\approx 1$, as $M \ll 1$.} Thus, the steady-state solution is 
\bea
P(v_x,v_y,\h) =  {\frac{M}{2\pi}} e^{-\frac{M}{2} [(v_x - \l \cos\h)^2 + (v_y - \l \sin\h)^2]}.
\eea
The velocity distribution $P(v_x,v_y)$ can be obtained by integrating over all possible $\h$, $P(v_x,v_y) =(1/2\pi)\int_0^{2\pi} d\h P(v_x,v_y,\h)$. The marginal distribution,
\bea \label{eq_prob1}
P(v_x) = \int_{0}^{2\pi} \frac{d \h}{2\pi}~ \sqrt{\frac{M}{2\pi}} e^{-\frac{M}{2}(v_x - \l \cos\h)^2},
\eea
by symmetry has the same form as $P(v_y)=\int_{0}^{2\pi} \frac{d \h}{2\pi}~ \sqrt{\frac{M}{2\pi}} e^{-\frac{M}{2}(v_y - \l \sin\h)^2}$. As shown in Fig.\ref{fig_pvx}($a$), this distribution already describes the small inertia crossover between the passive Gaussian to active non-Gaussian (bimodal symmetric) nature, capturing the numerical simulation results at $M=10^{-4}, 10^{-2}$ for $\l=100$. Thus, the previously proposed perturbative calculation~\cite{herrera2021maxwell} is not necessary to qualitatively describe the non-Gaussian feature. Moreover, as the perturbative calculation is valid only for small $M$ limits, it cannot capture the large $M$ re-entrance to Gaussian behavior. 

At large $M \gg 1$, orientational relaxation leads to effective equilibrium-like Gaussian distributions with {%\color{blue} 
fluctuations governed by the  kinetic temperature $\kb T_{\rm kin} \approx 1+\l^2/[d(d-1)] $~(see Eq.(\ref{eq_Tkin}))} 
\bea \label{eq_prob2}
P(v_x, v_y) = {\frac{M}{2\pi \kb T_{\rm kin}}} e^{-\frac{M}{2 \kb T_{\rm kin}}(v_x^2+v_y^2)},~
P(v_i)=\sqrt{\frac{M}{2 \pi \kb T_{\rm kin}}}e^{\frac{-Mv_i^2}{2\kb T_{\rm kin}}}. 
\eea 
In the last expression, $v_i$ denotes both $v_x$ and $v_y$. 
This expression captures the simulation results at $M=10^2$ and $\l=100$, thereby describing the crossover to passive Gaussian behavior at large enough $M$. Thus, equations~(\ref{eq_prob1}) and (\ref{eq_prob2}) together capture the re-entrance observed as a function of $M$ as evidenced from the numerical simulations shown in Fig.\ref{fig_pvx}.

Similar arguments as above can be used to describe the probability distribution of  $v_\parallel$ and $v_\perp$ in the two limits of small and high inertia. At small $M$ their distributions take the form
\bea
\label{eq_vsm} P(v_\parallel) = \sqrt{\frac{M}{2\pi}} e^{-\frac{M}{2}(v_\parallel - \l)^2}, ~
P(v_\perp) =  \sqrt{\frac{M}{2\pi}} e^{-\frac{M}{2}v_\perp^2}.
\eea
These expressions agree with the numerical result at $M = 10^{-4}$ (Fig.~\ref{fig_pvx}). Equation~(\ref{eq_vsm}) also suggests $\la \vpl \ra = \l$, $\la \vp \ra = 0$,  $\la \d \vpl^2\ra \sim 1/M$ and $\la \d \vp^2\ra \sim 1/M$ at small $M$, agreeing with equations~(\ref{eq_vplst}) and  (\ref{eq_dv2p}). 
In the large $M$ limit, distributions are described by the Pe-dependent kinetic temperature,
\bea
\label{eq_vlm} P(v_\parallel) = \sqrt{\frac{M}{2 \pi \kb T_{\rm kin}}} e^{-\frac{M}{2\kb T_{\rm kin}}v_{\parallel}^2},~ %\\
P(v_\perp) = \sqrt{\frac{M}{2 \pi \kb T_{\rm kin}}} e^{-\frac{M}{2\kb T_{\rm kin}}v_{\perp}^2} .
\eea
The above expressions are tested against the numerical result at $M = 10^{2}$ and show good agreements (see Fig.~\ref{fig_pvx}). Equation~(\ref{eq_vlm}) suggests $\la \vpl \ra=0=\la \vp \ra$ and the variance  $\la \d \vpl^2\ra = \la \d \vp^2\ra \sim 1/M$ at the large $M$ limit, agreeing with equations~(\ref{eq_vplst}) and (\ref{eq_dv2p}).

 %%----------------  mapping to harmonic trap -------------------------------%%
The velocity evolution under active drive Eq.(\ref{eom:speed_abp_speed_fluct}) can formally be mapped to the position evolution of an overdamped ABP in a harmonic trap $d r_i = (v_a u_i -\beta r_i)dt + \sqrt{2} \,dB_i(t)$~\cite{Chaudhuri2020}, with dimensionless activity $v_a$ and trap stiffness $\beta$. The mapping gives $M=\beta^{-1}$, $\l=v_a/\beta$ and the strength of the thermal noise $\beta^{-2}$. Thus, although the algebra is equivalent, extracting the physical meaning from such mapping requires care.  A further subtlety involves the coupling of positional evolution to velocity for inertial ABPs, which we explore next. 

 %%----------------  Displacement -------------------------------%%
\section{Displacement moments, active diffusivity, and swim pressure}
\label{sec_disp}
In this section, we calculate moments of displacement vector to explore their time evolution. We begin by using $\psi = \rv$ and the initial condition $\la \psi \ra_0 = 0$ in equation(\ref{observable}) to get $\la \rv \ra_s = \la \vv \ra_s/s $ . Further using equation(\ref{v_lap}), we get
\bea \label{eqrs}
\la \rv \ra_s = \f{\vv_0}{s (s+1/M)} + \f{(\l /M) \uv_0}{s(s+1/M)(s+d-1)}. 
\eea
Its inverse Laplace transform gives
\bea
\fl
    \la \rv \ra (t) = M\vv_0 +\frac{\l \uv_0}{(d-1)} + \frac{\l \uv_0\, e^{-(d-1)t}}{(d-1)[(d-1)M-1]}  
     - \left( \vv_0 + \frac{\l \uv_0}{(d-1)M-1} \right) M \, e^{-t/M}
\eea
In the limit of $M\to 0$ the above expression gives $\la \rv (t) \ra = \f{\l \uv_0}{(d-1)} [1-e^{-(d-1)t}]$ in agreement with the evolution of overdamped ABPs~\cite{Shee2020}.  

%%----------------  Mean - Squared Displacement -------------------------------%%

 \subsection{Mean squared displacement (MSD) and asymptotic diffusivity}
\begin{figure}[t]
    \centering
    \includegraphics[scale = 0.8]{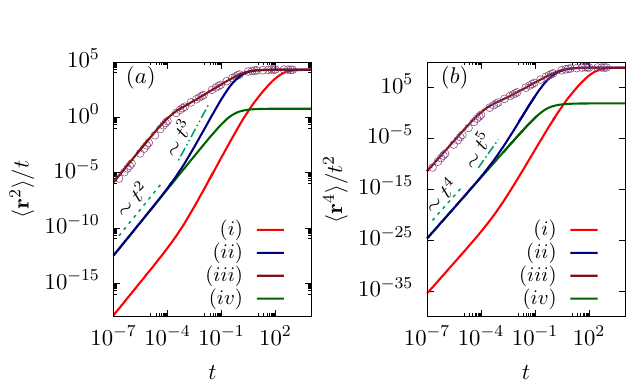}
    \caption{Evolution of $\la \mathbf{r^2} \ra/t$ and $\la \mathbf{r}^4 \ra/t^2$ with time in $d = 2$ with initial velocity $\vv_0 = (0,0)$. Lines in ($a$) show plots of $\la \mathbf{r}^2 \ra$ obtained from equation (\ref{eqr2t}) and in ($b$) show plots of $\la \mathbf{r}^4 \ra$ obtained from solving equations (\ref{r4s})-(\ref{uvrvs}) and taking inverse Laplace transform. The points $(i)$, $(ii)$, $(iii)$, and $(iv)$ are the same as shown in Fig.\ref{fig_vdistribution}($b$). The points denote simulation results.}
    \label{fig_r2_r4}
\end{figure}

Let us now consider $\psi  = \rv^2$ and the initial condition $\rv(t=0)=0$ in equation(\ref{observable}). It is easy to see $\la \vv \cdot \nabla \psi \ra_s = 2\la \vv \cdot \rv \ra_s$, $\vv \cdot \nabla_v \psi = 0$, $ \uv \cdot \nabla_v \psi = 0$, $ \nabla_v^2 \psi  = 0$, and $\nabla_u^2 \psi = 0$. Thus, equation(\ref{observable}) leads to 
$\la \rv^2 \ra_s = 2\la \vv \cdot \rv \ra_s /s$. 
The calculation of the position-velocity cross-correlation $\la \vv \cdot \rv \ra_s $ proceeds in the same way using $\psi = \vv \cdot \rv$ in equation(\ref{observable}), using initial condition $\la \psi \ra_0 = 0 $, $ \la \vv \cdot \nabla \psi \ra_s = \la \vv^2 \ra_s $, $ \la \vv \cdot \nabla_v \psi \ra_s = \la \vv \cdot \rv \ra_s $, $\la \uv \cdot \nabla_v \psi \ra_s = \la \uv \cdot \rv \ra_s$, $\nabla_v^2 \psi = 0$, and $\nabla_u^2 \psi = 0$. This leads to 
\bea \label{eqvrs}
(s + 1/M)\la \vv \cdot \rv \ra_s = \la \vv^2 \ra_s + \l \la \uv \cdot \rv \ra_s/M. 
\eea
To complete the calculation, one needs to evaluate the position-orientation cross-correlation $\la \uv \cdot \rv \ra_s$, using $\psi = \uv \cdot \rv$ in equation(\ref{observable}) with $\la \psi \ra_s = 0$, $\la \vv \cdot \nabla \psi \ra_s = \la \uv \cdot \vv \ra_s $, $\vv \cdot \nabla_v \psi  = 0 $, $ \uv \cdot \nabla_v \psi = 0$, $\nabla_v^2 \psi = 0$, and $ \la \nabla_u^2 \psi  \ra_s= -(d-1) \la \uv \cdot \rv \ra_s$ to obtain 
\bea \label{equrs}
(s  + (d-1)) \la \uv \cdot \rv \ra_s = \la \uv \cdot \vv \ra_s. 
\eea
Note that $\la \vv^2 \ra_s $ and $\la \uv \cdot \vv \ra_s $ were already calculated in equations (\ref{finalv2s}) and (\ref{eq_uvs}). 
Thus, we find 
\bea \label {eqfinalr2s}
    \fl \la \rv^2 \ra_s =  \frac{2}{s(s+1/M)} \left[ \frac{1}{s+2/M} \left ( \vv_0^2 + \frac{2d}{s M^2} \right. \right.{\left. \left. +\frac{2\l\{(\uv_0 \cdot \vv_0)sM + \l\}}{sM\{1+[s+(d-1)]M\}} \right) \right.} \nn\\
    {\left. + \frac{\l \{(\uv_0 \cdot \vv_0)sM + \l\}}{sM(s+(d-1))\{1+(s+(d-1))M\}} \right].}
\eea
The inverse Laplace transform of equation(\ref{eqfinalr2s}) gives the full time dependence of MSD,
\bea \label {eqr2t}
\fl
\la \rv^2 \ra (t)  = 2 t \left(d +\frac{\l ^2}{d-1}\right) +  \frac{2 \l [(\uv_0 \cdot\vv_0) (d-1) M-\l ]}{(d-1)^2 [(d-1) M-1]}  e^{-(d-1)t} \nn\\
\fl -\frac{2 \Pe  M  \{(\uv_0 \cdot\vv_0) [(d-1) M+1]-\Pe \}}{(d-1)^3 M^2-d+1}e^{-((d-1)+M^{-1})t}\nn\\
 \fl + M\left( -(d-\vv_0^2M) +\frac{2(\uv_0 \cdot \vv_0)M\l - M\l^2}{(d-1)M-1} \right)e^{-2t/M}\nn\\
\fl +2M \left[ (2d-\vv_0^2M) + \frac{(\uv_0 \cdot \vv_0)(1-2(d-1)M)\l}{(d-1)((d-1)M-1)}  + \frac{\{2(d-1)M-1\}\l^2}{(d-1)((d-1)M-1)} \right] e^{-t/M}  \nn\\
\fl + M (\vv_0^2 M-3 d) +\frac{2(\uv_0 \cdot \vv_0) \Pe  M}{d-1} -\frac{\Pe ^2 \{(d-1) M (3 (d-1) M+4)+2\}}{(d-1)^3 M+(d-1)^2}
\eea
As shown in Fig.~\ref{fig_r2_r4}($a$), the above equation captures the numerical simulation results of MSD. Moreover, the MSD shows a curious $\sim t^3$ scaling at short times before reaching the asymptotic diffusive scaling. Note that the long-time diffusive behavior  
\bea \label{eq_Deff}
\lim_{t \to \infty} \la \rv^2 \ra (t) &=  2d D_{\rm eff }t~ {~\rm with~~} D_{\rm eff} = 1+\l^2/(d(d-1))
\eea
is independent of inertia, and the diffusion constant has the same form as overdamped ABPs~\cite{Shee2020, Scholz2018}. 
To analyze the short-time behavior, we expand $ \la \rv^2 \ra (t)$ around $t = 0$
 \bea \label{eq_r2tseries} 
 \la \rv^2 \ra (t)  = \vv_0^2 t^2 + {\cal A}_2 t^3 + {\cal B}_2 t^4 +{\cal O}\left(t^5\right)   
 \eea
 where ${\cal A}_2=(-3 \vv_0^2 M+3 (\uv_0 \cdot \vv_0) \Pe  M+2 d)/(3 M^2)$
 and ${\cal B}_2 = [7 \vv_0^2 M-2 (\uv_0 \cdot \vv_0) \Pe  M (2 (d-1) M+5)-6 d+3 \Pe ^2 M]/(12 M^3)$. 
Equation\,(\ref{eq_r2tseries}) agrees with the observation of short-time ballistic behavior of MSD~\cite{Scholz2018} if the initial velocity is drawn from the equilibrium distribution $\vv_0^2 = d/M$. 
However, a choice of $\vv_0 = {\bf 0}$ leads to $\la \rv^2 \ra \sim t^3$ at short times in agreement with Fig.\ref{fig_r2_r4}, and the series expansion simplifies to
 \bea 
  \fl
   \la \rv^2 \ra (t)  = \frac{2 d }{3 M^2}t^3+\frac{ \left(\l^2 M-2 d\right)}{4 M^3}t^4 -\frac{ \left(2 \l^2 M^2(d-1)+5 \l^2 M-7 d\right)}{30 M^4}t^5 +{\cal O}\left(t^6\right).  
 \eea
The expansion explains the short-time crossover from $\la \rv^2 \ra\sim t^3$ to $\la \rv^2 \ra\sim t^4$ observed from numerical simulations near $t_I = 8dM/[3(\l^2M-2d)]$ provided $\l^2M > 2d$, a criterion fulfilled by the lines ($i$) ($M=100,\l=100$) and ($ii$) ($M=0.2, \l=100$) in Fig.~\ref{fig_r2_r4}($a$). Note that for lines ($iii$) ($M=10^{-4}$, $\l=100$) and ($iv$) ($M=0.2$, $\l=1$), the criterion $\l^2M > 2d$ is not obeyed and as a result beyond the initial $\sim t^3$ scaling,  $\la \rv^2 \ra$ directly approaches the asymptotic $\sim t$ behavior. 

%---------------------- Figure - 3----------------------%
%
%------------------------------------------------------%

\subsection{Position orientation cross-correlation and swim pressure}

While calculating $\la \rv^2\ra$, we computed the cross-correlation $\la \uv \cdot \rv \ra$. Here, we consider its role in determining the swim pressure of ABPs. Note that the virial term for an ideal gas of $N$ ABPs due to activity $\pi^a = (1/d\,V)\sum_{i=1}^N \la \rv'_i \cdot \g \l' \uv_i  \ra$ with $V$ denoting an effective volume~\cite{Bialke2015, Takatori2014a, Mallory2013a}. The dimensionless pressure has the form  ${\cal P} = \r (\l/d) \la \uv \cdot \rv \ra_{st}$, where $\r$ denotes the dimensionless particle density and 
$\la \uv \cdot \rv\ra_{st} = \f{\l}{(d-1)(1+(d-1)M)}$. 
Thus, the swim pressure
\bea
{\cal P} = \r \f{\l^2}{d(d-1)} \f{1}{[1+(d-1)M]}.  
\eea
Unlike in passive particles or overdamped ABPs, the swim pressure turns out to be a function of inertia $M$.
Note that ${\cal P}$ can be re-expressed in terms of the excess diffusivity due to activity $D^{(a)}_{\rm eff} = \f{\l^2}{d(d-1)}$ such that ${\cal P} = \r D^{(a)}_{\rm eff} \, \f{1}{[1+(d-1)M]}$. In the limit of $M\to 0$, swim pressure is the maximum, ${\cal P} = \r D^{(a)}_{\rm eff}$. This result in the overdamped limit agrees with earlier calculations for ABPs in $d=3$~\cite{Takatori2014a}. However, for inertial ABPs, the swim pressure ${\cal P}$ vanishes as $M^{-1}$ in the limit of large inertia.   The decrease of swim stress with translational inertia was observed before in Ref.~\cite{Takatori2017, sandoval2020pressure}.

%%----------------  Fourth order moment of displacement -------------------------------%%
\subsection{Fourth moment of displacement}
The calculation of the fourth moment in the Laplace space is long but straightforward. We show the essential steps of this calculation in \ref{sec_r4}.  
The inverse Laplace transform can be used to obtain $\la \rv^4 \ra(t)$. As the expression is too lengthy to show here, we plot the expression $\la \rv^4 \ra(t)$ as a function of time and compare it with the result of simulation for various $M$ and $\l$ values in Fig.\ref{fig_r2_r4}($b$) to find good agreement. Here, we analyze the expression in the short and long time limits. 
The asymptotic behavior of $\la {\rv}^4 \ra$ in the long time limit gives 
\bea
\lim_{t\to\infty} \la {\rv}^4 \ra ({t}) = \frac{4 (d+2) \left(d(d-1) +\l ^2\right)^2}{d(d-1)^2 } {t}^2,
\eea
with the coefficient of the $t^2$ scaling independent of inertia. The $t^6$ scaling in the $t\to 0$ limit observed in Fig.\ref{fig_r2_r4}($b$) can be understood performing a series expansion of the expression around $t=0$ with initial velocity $\vv_0 = {\bf 0}$ 
\bea 
\la \rv^4 \ra (t) = \frac{4d(d+2)}{9M^4}t^6 + \frac{(d+2)(\l^2M-2d)}{3M^5}t^7 
+ {\cal C}_3 t^8 + {\cal {O}} (t^9),
\eea
where ${\cal C}_3=\{\l^2 M^2 [45 \l^2-64 (d^2+d-2)]-340 (d+2) \l^2 M+404 d (d+2)\}/(720 M^6)$. 
Note that the shortest time $\sim t^6$ scaling is due to equilibrium fluctuations, independent of activity.
This behavior with initial velocity $\vv_0={\bf 0}$ starkly contrasts the shortest-time $\la \rv^4 \ra (t)\sim t^2$ scaling observed for overdamped ABPs~\cite{Shee2020}. The short-time $\sim t^2$ scaling for overdamped ABPs is recovered if one keeps a non-zero $\vv_0$ while taking the limit $M\to 0$.    
The first non-equilibrium influence comes in $\la \rv^4 \ra(t)$ determining the crossover to $\sim t^7$ scaling at $t_I = 4dM/[3(\l^2M-2d)]$. This crossover will be observable if $\l^2 M > 2d$. As a result, we see this crossover in lines ($i$) and ($ii$) of Fig.\ref{fig_r2_r4}($b$), but not in lines ($iii$) and ($iv$). 

{%\color{blue} 
\section{Discussion: Comparison with RTP and AOUP}
\label{sec_discuss}
As the introduction mentions, the ABP model is closely related to the RTP and AOUP models. They differ only by the statistical properties of the active forces.  However, up to the second moment, these properties are equivalent. The averages over the active forces over a time scale longer than the correlation time vanish. The two-time correlations for the active forces show exponential decay with characteristic correlation times. As a result, our predictions for ABPs up to second-order moments remain valid for these other two models and, with suitable mapping, can be used to predict exact results for them. 

To illustrate this, let us denote the position and velocity vectors $(\rv, \vv)$ to express the dynamics of an active particle in $d$-dimensions, %in two dimensions,
\bea
\dot \rv &=& \vv, \nn\\
m \dot \vv &=& -\g (\vv - \vv_a) + \g \sqrt{2 D}\, {\boldsymbol \eta}(t), \nn
\eea
where the only difference between the three models is in the dynamics of the active force $\g \vv_a$. ABPs differ from RTPs in their rotational relaxation. 
For the orientational diffusion of ABPs, the heading direction auto-correlation $\la \uv (t) \cdot \uv (t') \ra = e^{-(d-1)D_r |t-t'|}$ using equation\,(\ref{observable}).
RTPs undergo discrete tumbling events with rate $\g_r$~\cite{Cates2013, Solon2015b, Santra2020a}, such that the orientation propagator 
$P(\uv,t| \uv_0,0) = e^{-\g_r t} \d(\uv-\uv_0) + (1-e^{-\g_r t}) \f{1}{\Omega}$ with $\Omega$ the area of a $d$-dimensional unit sphere.
Thus, the orientational correlation controlling the active forces follows 
$\la \uv(t)\cdot \uv(t') \ra = e^{-\g_r|t-t'|}$. 
In contrast, within the AOUP model, 
using $\t_a \dot \vv_a = -\vv_a + v_{ao}\sqrt{2\t_a/d}~ {\boldsymbol \eta_v}(t)$
in $d$ dimensions 
with Gaussian white noise ${\boldsymbol \eta_v}(t)$~\cite{Caprini2021, Shee2020},  the steady-state auto-correlation follows
$\la \vv_a(t) \cdot \vv_a(t') \ra = v_{ao}^2 e^{-|t-t'|/\t_a}$.
Thus, the active force in the three models is characterized by $\la \vv_a \ra=0$ and   
\begin{equation}
  \la \vv_a (t) \cdot \vv_a(t') \ra = \left\{\begin{array}{r@{}l@{\qquad}l}
   v_a^2 \, e^{-(d-1) D_r |t-t'| } &  & \textrm{for \, \, ABP,} \\[\jot]
   v_r^2 \, e^{-\g_r |t-t'|}  &  &\textrm{for \, \, RTP,} \\[\jot]
      v_{ao}^2 e^{-|t-t'|/\t_a} &  & \textrm{for \, \, AOUP}.
  \end{array}\right.
\end{equation}
In the above relations, we used $v_r$ and $v_a$ to denote the amplitude of active velocity in the RTP and ABP models, respectively. Up to the second moment, all the models are equivalent with the mapping $ v_a = v_r = v_{ao}$ and $(d-1) D_r=\g_r = 1/\t_a$. The connection between ABP and RTP with $\g_r \leftrightarrow (d-1)D_r$ was noted before in Ref.\cite{Cates2013}. 

The units of time and length are set in the three models by (i)~$\t_r=D_r^{-1},\, \ell=\sqrt{D/D_r}$~(ABP), 
(ii)~$\t_r=\g_r^{-1},\, \ell=\sqrt{D/\g_r}$~(RTP),
(iii)~$\t_r=\t_a, \, \ell=\sqrt{D \t_a}$~(AOUP). Using these mappings and the definitions $M=m/\g \t_r$, Pe$=v_a/\sqrt{D/\t_r}$ the expressions derived for  $\la \vv \ra$, $\la \vv^2\ra$, $\la \rv \ra$, $\la \rv^2\ra$ within the ABP model remains valid for RTP and AOUP models. The same is true for the expressions of kinetic temperature, diffusivity, and pressure. 
The ensuing results agree with, e.g., the earlier AOUP calculations of Ref.\cite{Caprini2021} in $d=2$.  
Finally, the velocity distribution in RTP can show Gaussian-non-Gaussian crossovers, as in ABP. In contrast, given the inbuilt Gaussian nature, AOUP can not give rise to any non-Gaussian distribution, and all its higher-order dynamical moments can be calculated using the first two moments. Whether experimental observations demonstrate a non-Gaussian nature or not can be used to determine the potential application of the AOUP model in their analysis.
}

\section{Conclusion}
\label{sec_conc}
In this paper, we conducted a thorough analysis of the dynamics of free inertial Active Brownian Particles (ABP) in the presence of thermal noise. 
To achieve this, we employed a method based on the Laplace transform of the Fokker-Planck equation, allowing us to obtain expressions of time-dependent dynamical moments of arbitrary observables in any $d$-dimensional space. We obtained explicit expressions for several moments of velocity and displacement and equal time cross-correlations. The analytical predictions were rigorously validated through direct numerical simulations in two dimensions.

Our study has uncovered several intriguing characteristics of inertial active particles, setting them apart from their overdamped and passive counterparts. Notably, the inertia-dependent characteristics manifested in the steady-state kinetic temperature and swim pressure while leaving the late-time diffusivity unchanged. Specifically, the kinetic temperature initially increased before saturation, while the pressure consistently demonstrated a monotonic decrease. 

Furthermore, our study revealed fascinating re-entrant crossovers in the steady-state velocity distribution, transitioning from ``passive" Gaussian to ``active" non-Gaussian behaviors. To illustrate this phenomenon, we constructed a detailed ``phase diagram" in the activity-inertia plane, utilizing the exact expression of $d$-dimensional kurtosis.
Additionally, we proposed approximate velocity distribution formulas that are closely aligned with numerical simulations, especially in small and large inertia limits. These formulations also captured the transition from low inertia passive to active states.
{%\color{blue} 
Moreover, our findings can be expanded to encompass related models such as RTP and AOUP, enabling predictions up to second moments of dynamic variables through suitable mappings.} 
 
Our exact expressions, beyond describing steady states, could be used to explain time-dependent crossovers observed in various moments of velocity and displacement. Overall, this study provides significant insight into the intricate interplay between activity, inertia, and thermal noise governing the behavior of inertial active particles.          

\section*{Acknowledgments}
DC thanks Abhishek Dhar and Fernando Peruani for collaborations on related topics, acknowledges research grants from DAE (1603/2/2020/IoP/R\&D-II/150288) and SERB, India (MTR/2019/000750), and thanks ICTS-TIFR, Bangalore, for an Associateship and for hosting in the program -  Soft and Living Matter: from Fundamental Concepts to New Material Design (code: ICTS/slm2023/8).

%\vskip .2cm
%\noindent
%{\bf Funding information:}
%DC acknowledges research grants from DAE (1603/2/2020/IoP/R\&D-II/150288) and SERB, India (MTR/2019/000750).

\appendix

%%----------------  Fourth order moment of velocity -------------------------------%%

 \section{Fourth moment of velocity}
\label{sec_v4}
Inserting equations (\ref{uvv2s}),  (\ref{eq_uvs}), (\ref{finalv2s}) and (\ref{eq_vp})
into equation (\ref{v4s}), we get an exact form of the fourth order moment of velocity $\la \vv^4 \ra_s$ in Laplace space 
\small \bea \label{eq_v4fs}
\fl
\la \vv^4 \ra_s = \frac{8d(d+2)}{M^4s(s+2/M)(s+4/M)}+\frac{8(d+2)\l^2}{M^4s(s+2/M)(s+4/M)(s+1/M+d-1)} \nn\\
\fl+\frac{8(d+2)\l^2}{M^4s(s+4/M)(s+3/M+d-1)(s+1/M+d-1)} \nn\\
\fl+\frac{8 d\l^2}{M^4s(s+2/M)(s+4/M)(s+3/M+d-1)} +\frac{16 \l^2}{M^4s(s+2/M+2d)(s+3/M+d-1)(s+4/M)} \nn\\
\fl+\frac{32 d \l^2}{M^4s(s+2/M)(s+4/M)(s+2/M+2d)(s+3/M+d-1)} \nn\\
\fl+\frac{8\l^4}{M^4s(s+2/M)(s+4/M)(s+1/M+d-1)(s+3/M+d-1)} \nn\\
\fl+\frac{16\l^4}{M^4s(s+4/M)(s+1/M+d-1)(s+3/M+d-1)(s+2/M+2d)} \nn\\
\fl+ \frac{32\l^4}{M^4s(s+2/M)(s+4/M)(s+1/M+d-1)(s+3/M+d-1)(s+2/M+2d)} \nn\\
\fl+\frac{\vv_0^4}{s+4/M} + \frac{8(\uv_0\cdot \vv_0)^2\l^2}{M^2(s+4/M)(s+3/M+d-1)(s+2/M+2d)}+\frac{4(d+2)\vv_0^2}{M^2(s+2/M)(s+4/M)} \nn\\
\fl +\frac{4\l^2\vv_0^2}{M^2(s+2/M)(s+4/M)(s+3/M+d-1)} \nn\\
\fl+\frac{16\l^2\vv_0^2}{M^2(s+2/M)(s+4/M)(s+2/M+2d)(s+3/M+d-1)} \nn\\
\fl+ \frac{8(d+2)\l (\uv_0 \cdot \vv_0)}{M^3(s+2/M)(s+4/M)(s+1/M+d-1)} +\frac{8(d+2)\l (\uv_0 \cdot \vv_0)}{M^3(s+4/M)(s+1/M+d-1)(s+3/M+d-1)} \nn\\
\fl+ \frac{8\l^3 (\uv_0 \cdot\vv_0)}{M^3(s+2/M)(s+4/M)(s+1/M+d-1)(s+3/M+d-1)} \nn\\
\fl+\frac{16 \l^3 (\uv_0 \cdot \vv_0)}{M^3(s+4/M)(s+1/M+d-1)(s+3/M+d-1)(s+2/M+2d)} \nn\\
\fl+\frac{32 \l^3 (\uv_0 \cdot \vv_0)}{M^3(s+2/M)(s+4/M)(s+1/M+d-1)(s+3/M+d-1)(s+2/M+2d)} \nn\\
\fl+ \frac{4(\uv_0 \cdot \vv_0) \vv_0^2}{M(s+4/M)(s+3/M+d-1)}
\eea
The inverse Laplace transform of equation (\ref{eq_v4fs}) gives an exact form of $\la \vv^4 \ra$(t)
 \small \bea \label{eq_v4t}
\fl
 \la \vv^4 (t) \ra = \frac{d(d+2)}{M^2} + \frac{2(d+2)\l^2}{M+(d-1)M^2}+\frac{[(d+2)M+3]\l^4}{(dM+1)[(d-1)M+1][(d-1)M+3]} \nn \\
\fl + \frac{2(d+2)\{d^2-[d(d-1)+\l^2]^2M^2\}}{M^2d[(d-1)^2M^2-1]}e^{-2t/M} \nn\\
 \fl +\frac{e^{-4t/M}}{M^2[(d-1)M-3][(d-1)M-1](dM-1)}  \left[\{(d+2)M-3\}M^2 \l^4 \right. + \nn\\
  \fl 2(d+2)((d-1)M-3)(dM-1)M\l^2  + 
\left. d(d+2)\{(d-1)M-3\}\{(d-1)M-1\}(dM-1) \right] \nn\\
  \fl + \frac{4(d-1)\l^4}{d(dM-1)(dM+1)[(d+1)M-1][(d+1)M+1]} e^{-2(d+M^{-1})t} \nn\\
  \fl + \frac{4\l^2 \{(d+2)[(d-1)M-3][(d+1)M+1] + \l^2 [(d-7)M^2-3M]\}}{M[(d-1)M-3][(d-1)M-1][(d-1)M+1][(d+1)M+1]} e^{-((d-1)+M^{-1})t} \nn\\
  \fl - \frac{4\l^2 \{(d+2)[(d-1)M+3][(d+1)M-1] +\l^2 [(d-7)M^2+3M]\}}{M[(d-1)M-1][(d-1)M+1][(d-1)M+3][(d+1)M-1]} e^{-((d-1)+3M^{-1})t}\nn\\
  \fl  + \vv_0^4 e^{-4t/M} + \frac{4\l (\uv_0 \cdot \vv_0)\vv_0^2}{((d-1)M-1)} \left[ e^{-4t/M} -e^{-((d-1)+3M^{-1})t} \right] \nn\\
  \fl  + \frac{4(\uv_0 \cdot \vv_0)^2 \l^2
}{(dM-1)(dM+M-1)((d-1)M-1)} \left[ ((d-1)M-1) e^{-2(d+M^{-1})t} \right. \nn\\
  \fl \left. -2(dM-1)e^{-(d-1+3M^{-1})t}+(dM+M-1)e^{-4t/M}  \right] \nn\\
  \fl  + \frac{2(d+2) \vv_0^2}{M} \left( e^{-2t/M} - e^{-4t/M} \right) + 2\l^2 \vv_0^2 \left(- \frac{2e^{-2(d+M^{-1})t}}{d(dM-1)(dM+M-1)} +\frac{d+2}{d((d-1)M+1)}e^{-2t/M}   \right. \nn\\
\fl \left. + \frac{1-(d+2)M}{(dM-1)((d-1)M-1)} e^{-4t/M}+ \frac{2(d+5)M-1}{(dM+M-1)((d-1)M+1)((d-1)M-1)} e^{-(d-1+3M^{-1})t} \right) \nn\\ 
\fl  +\l (\uv_0 \cdot \vv_0) \frac{4 (d+2) \left(e^{2t/M}-1\right) e^{-\left(d+4M^{-1}\right)t} \left(e^{d t}-e^{(1+M^{-1})t}\right)}{M ((d-1) M-1)} \nn\\
\fl  + 4\l^3 (\uv_0 \cdot \vv_0) \left(  \frac{((d-7) M+3) e^{-(d-1+3M^{-1})t}}{((d-1) M-1) ((d-1) M+1) (d M+M-1)}  \right. 
\nn\\
\fl \left. -\frac{((d-7)M-3 ) e^{-((d-1)+M^{-1})t}}{((d-1) M-3) ((d-1) M-1) (d M+M+1)}   -\frac{2 (d-1) e^{-2 \left(d+M^{-1}\right)t}}{d (d M-1) (d M+M-1) (d M+M+1)} \right. \nn\\
\fl  \left. +\frac{(d+2) e^{-2 t/M}}{d \left((d-1)^2 M^2-1\right)} 
+ \frac{(3-(d+2) M) e^{-{4 t}/{M}}}{((d-1) M-3) ((d-1) M-1) (d M-1)} \right)
\eea

\section{Fourth moment of displacement}
\label{sec_r4}
The calculation of $\la \rv^4 \ra_s$ requires following steps :
\bea
\label{r4s}\fl s \la \rv^4 \ra_s = 4 \la (\vv \cdot \rv)\rv^2 \ra_s \\ 
\fl (s+1/M) \la (\vv \cdot \rv)\rv^2 \ra_s  = \la \vv^2 \rv^2 \ra_s + 2 \la (\vv \cdot \rv)^2 \ra_s + (\l/M) \la (\uv \cdot \rv)\rv^2 \ra_s \\
\fl (s+2/M) \la \vv^2 \rv^2 \ra_s = 2\la (\vv \cdot \rv)\vv^2 \ra_s + (2\l/M) \la (\uv \cdot \vv) \rv^2 \ra_s + (2d/M^2)\la\rv^2 \ra_s \\
\fl (s+3/M) \la (\vv \cdot \rv)\vv^2 \ra_s = \la \vv^4 \ra_s + (\l/M) \la (\uv \cdot \rv)\vv^2 \ra_s +(2\l/M)\la (\uv \cdot \vv)(\vv \cdot \rv)\ra_s \nn\\
\fl + 2((d+2)/M^2)\la \vv \cdot \rv\ra_s \\
\fl (s+2/M +(d-1)) \la (\uv \cdot \vv)(\vv \cdot \rv) \ra_s = \la (\uv \cdot \vv) \vv^2 \ra_s +(\l/M)\la \vv \cdot \rv \ra_s \nn\\
\fl +(\l/M)\la (\uv \cdot \vv)(\uv \cdot \rv)\ra_s + (2/M^2) \la \uv \cdot \rv \ra_s \\
\fl (s+2/M +(d-1)) \la (\uv \cdot \rv)\vv^2 \ra_s = \la (\uv \cdot \vv)\vv^2 \ra_s + (2\l/M)\la (\uv \cdot\vv)(\uv \cdot \rv)\ra_s  \nn\\
\fl + (2d/M^2) \la \uv \cdot \rv \ra_s \\
\fl (s+1/M + 2d) \la (\uv \cdot \vv)(\uv \cdot \rv)\ra_s = \la (\uv \cdot \vv)^2 \ra_s + (\l/M) \la \uv \cdot \rv \ra_s + 2 \la \vv \cdot \rv \ra_s \\
\fl (s+1/M+(d-1)) \la (\uv \cdot \vv)\rv^2 \ra_s = 2 \la (\uv \cdot \vv)(\vv \cdot \rv) \ra_s + (\l/M) \la \rv^2 \ra_s \\
\fl (s+2/M) \la (\vv \cdot \rv)^2 \ra_s = 2 \la (\vv \cdot \rv)\vv^2 \ra_s +(2\l/M) \la (\uv \cdot \rv)(\vv \cdot \rv)\ra_s  +(2/M^2) \la \rv^2 \ra_s \\
\fl (s+(d-1)) \la (\uv \cdot \rv) \rv^2 \ra_s = \la (\uv \cdot \vv) \rv^2 \ra_s +2 \la (\uv \cdot \rv)(\vv \cdot \rv) \ra_s\\
\fl (s+1/M+(d-1)) \la (\uv \cdot \rv)(\vv \cdot \rv)\ra_s = \la (\uv \cdot \vv)(\vv \cdot \rv)\ra_s +\la (\uv \cdot \rv)\vv^2 \ra_s   \nn\\
\fl +(\l/M) \la (\uv \cdot \rv)^2 \ra_s\\
\fl (s+2d) \la (\uv \cdot \rv)^2 \ra_s = 2 \la (\uv \cdot \vv)(\uv \cdot \rv)\ra_s +2 \la \rv^2 \ra_s \label{uvrvs}
\eea
where $\la \vv^2 \ra_s $, $\la \vv^4 \ra_s $, $\la (\uv \cdot \vv)\vv^2 \ra_s $, $\la \uv \cdot \vv \ra_s $, $\la (\uv \cdot \vv)^2 \ra_s $, $\la \rv^2 \ra_s $, $\la \vv \cdot \rv \ra_s $, and $\la \uv \cdot \rv \ra_s $ were already calculated in equations (\ref{finalv2s}), (\ref{v4s}), (\ref{uvv2s}), (\ref{eq_uvs}), (\ref{eq_vp}), (\ref{eqfinalr2s}), (\ref{eqvrs}), and (\ref{equrs}) respectively. Inserting these back in equation(\ref{r4s}) one gets $\la \rv^4 \ra_s$.

\section{Velocity moments}
\label{sec_app_vel}
\begin{figure}[!t]
    \centering
    \includegraphics[scale = 0.75]{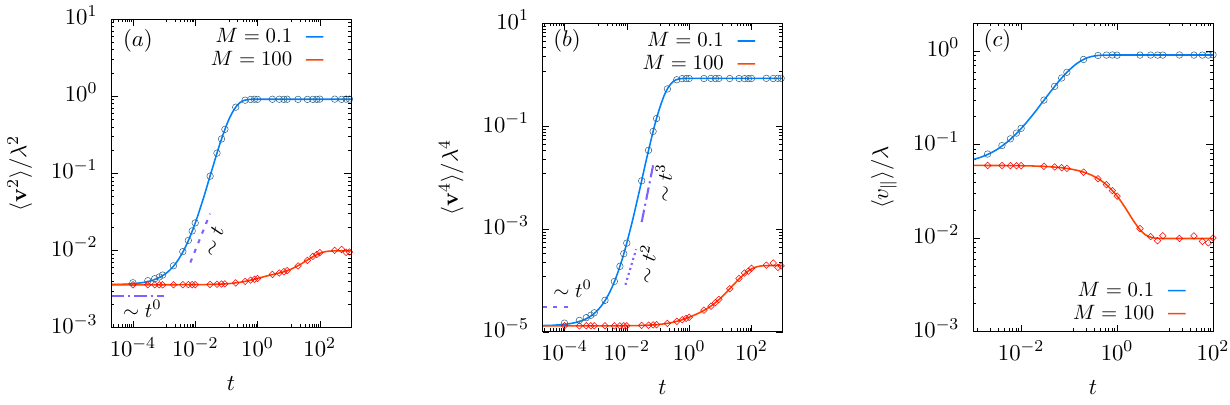}
    \caption{Evolution of moment of velocity in $d =2$ dimension with $\l = 100$, initial velocity $\vv_0 = (6,0)$ and initial orientation $\uv_0 = (1,0)$. Results for two $M$ values from simulations (points) and analytic calculations (lines) are shown.}
    \label{fig_v_initial}
\end{figure}

In this section, we supplement the results presented in the main text containing various velocity moments. In Fig.~\ref{fig_v_initial}, we show a comparison between analytic and numerical simulation results with initial conditions $\vv_0 = (6,0)$, $\uv_0 = (1,0)$. An expansion around $t=0$ using $\uv_0\cdot\vv_0 = v_0$ and $\vv_0^2 = v_0^2$ gives
\bea
\la \vv^2\ra(t)= v_0^2 + \frac{2[d+Mv_0(\l-v_0)]}{M^2}t + {\cal {C}}_4 t^2 + {\cal {O}} (t^3),
\eea 
and
\bea \label{eq_v4ap}
\la \vv^4 \ra(t)= v_0^4 + \frac{4 v_0^2[2+d+Mv_0(\l - v_0)]}{M^2} t + {\cal {C}}_5 t^2 + {\cal {O}} (t^3),
\eea
with $M^3{\cal {C}}_4 = \{\l^2M-2d+M \l v_0[-3-(d-1)M]+2Mv_0^2 $\}, $M^4 {\cal {C}}_5 = \{4d(d+2) -12(d+2)Mv_0^2 + 8M^2 v_0^4 + 2M \l v_0[4(d+2)+Mv_0^2(-7-(d-1)M)] +6M^2 \l^2 v_0^2 \}$. 
\begin{figure}[!h]
    \centering
    \includegraphics[scale = 0.7]{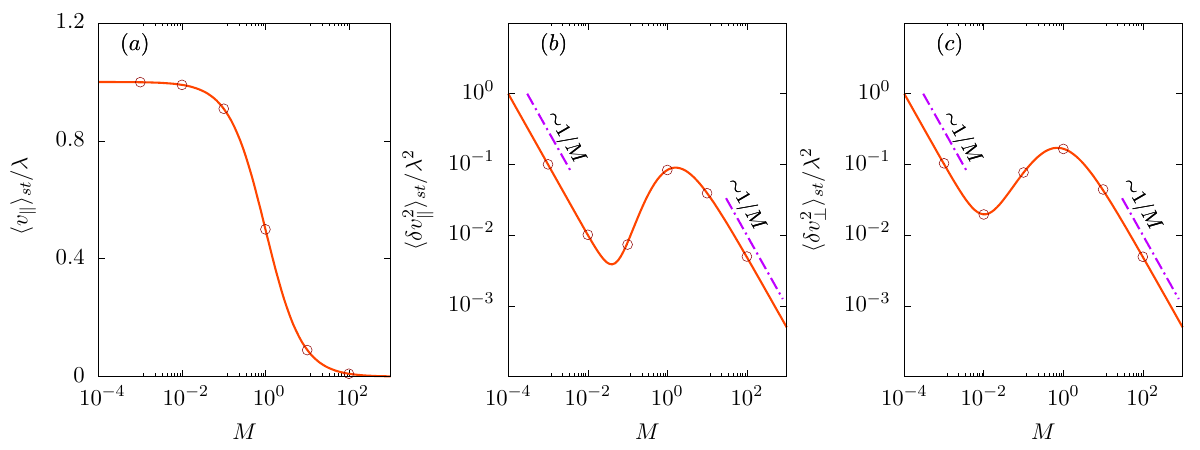} %{vcomponent2.pdf}
    \caption{Inertial dependence of velocity components: $\la \vpl \ra_{st}$, $\la \d \vpl^2\ra_{st}$ and $\la \d \vp^2\ra_{st}$ with $M$ at $\l=100$ and $d=2$. 
        The solid lines are the plots of equations (\ref{eq_vplst}), (\ref{eq_dv2p}), and (\ref{eq_vprp}) and the points denote simulation results.
        }
    \label{fig_vpl_st}
\end{figure}
These relations help to understand the differences in short-time scaling behaviors from the $\vv_0={\bf 0}$ case illustrated in the main text. Here, in the limit $t\to 0$, $\la \vv^2\ra = v_0^2$ and $\la \vv^4\ra = v_0^4$, as one would expect. The first $t$-dependence in $\la \vv^2\ra$ and $\la \vv^4\ra$ both show $\sim t$ scaling. Note the difference from $\vv_0={\bf 0}$ case particularly for $\la \vv^4\ra$, as the linear in $t$ term vanishes for $v_0=0$ in equation~(\ref{eq_v4ap}). Further, as shown in Fig.~\ref{fig_v_initial}, $\la \vpl\ra$ can even decrease with time before reaching the steady state.  

Fig.~\ref{fig_vpl_st} illustrates the steady-state variation of $\la \vpl \ra_{st}$, $\la \d \vpl^2\ra_{st}$ and $\la \d \vp^2\ra_{st}$ with $M$ in 2d using $\l=100$. The lines are analytic functions, and the points denote simulation results. 
As Fig.~\ref{fig_vpl_st}($a$) shows, $\la \vpl\ra_{st}$ decreases from $\l$ to $0$ monotonically, with increasing mass. Also, the $\la \d v_\parallel^2\ra_{st} \sim 1/M$ and $\la \d \vp^2\ra_{st} \sim 1/M$ scalings in the small and large $M$ limits are observable in Fig.~\ref{fig_vpl_st}.

\section*{References}

\bibliographystyle{unsrt} 

%\bibliography{../underdamped_abp}% Produces the bibliography via BibTeX.

\end{document}